\documentclass[%
 reprint,
superscriptaddress,
 amsmath,amssymb,
 aps,
]{revtex4-2}

\usepackage{graphicx}
\usepackage{dcolumn}
\usepackage{bm}
\usepackage{color}
\newcommand{\ket}[1]{|#1\rangle}
\newcommand{\bra}[1]{\langle #1 |}
\newcommand{\cra}[1]{\hat{#1}^\dagger}
\newcommand{\an}[1]{\hat{#1}}

\newcommand{\iner}[2]{\langle #1 | #2 \rangle}
\newcommand{\mat}[3]{\langle #1 |#2| #3 \rangle}

\newcommand{\av}[1]{\langle #1 \rangle}

\begin{document}

\preprint{APS/123-QED}

\title{Gauge Auxiliary-Field Quantum Monte Carlo Method for Many-Fermion Systems}

\author{Zhaozhan Zhang}
\email{zhangchzh6@mail.sysu.edu.cn}
\affiliation{Sino-French Institute of Nuclear Engineering and Technology, Sun Yat-sen University, Zhuhai, 519082, China}
\affiliation{Department of Physics, Graduate School of Science, The University of Tokyo, Tokyo, 113-0033, Japan}

\begin{abstract}
We propose novel Quantum Monte Carlo (QMC) methods for interacting many-fermion systems by leveraging the stochastic gauge freedom, originally developed in Gaussian phase-space QMC, within the phaseless auxiliary-field QMC (AFQMC) framework. In particular, we reinterpret the conventional force bias in phaseless AFQMC as a drift gauge and explore Fermi gauges based on natural orbitals of a reduced one-body density matrix defined via a mixed estimator, yielding stochastic, time-dependent Hartree-Fock-like dynamics. We propose a symmetry-projection sampling scheme to enhance the sampling efficiency. As a proof of concept, we apply these gauge-augmented AFQMC methods to a simple shell-model Hamiltonian: the Lipkin-Meshkov-Glick model. Numerical results illustrate the potential of stochastic gauges to enhance accuracy and reduce fluctuations, underscoring the promise for advancing these new techniques toward more realistic shell-model applications.
\end{abstract}

\maketitle
\section{Introduction}
Calculating interacting many-fermion systems remains one of the central challenges across diverse fields, from hadron and nuclear physics, condense matter physics, to quantum chemistry~\cite{hergert2020guided,anderson1972more,ruggenthaler2023understanding}. In nuclear physics, this challenge is further compounded by the complexity of the nuclear force and the strong correlations among a large number of nucleons, from just a few to over two hundred~\cite{TB:ring2004nuclear}. The nuclear shell model provides an excellent description of nuclear structure~\cite{caurier2005shell}, capturing exotic phenomena such as shape coexistence, phase transitions, and new magic numbers~\cite{otsuka2020evolution}. Unfortunately, conventional diagonalization methods suffer from combinatorial growth of the basis and particle number, especially near the dripline where large multi-shell spaces render the method infeasible. In addition, extending to heavy open-shell nuclei requires efficient treatment of pairing correlations~\cite{bohr1958possible}. 

In this context, quantum Monte Carlo (QMC) methods offer powerful nonperturbative alternatives~\cite{carlson2015quantum}. The Shell Model Monte Carlo (SMMC) method has been successfully applied to nuclear shell model~\cite{koonin1997shell,alhassid2017auxiliary}. However, with realistic effective interactions, the approach is plagued by the so-called fermion sign problem~\cite{baumgartner2013applications,nightingale1998quantum,loh1990sign}, which originates from the antisymmetric nature of fermionic wave function. This problem leads to an exponentially decaying signal-to-noise ratios as either system size or projection time increases, posing a fundamental challenge to scalability. Moreover, SMMC is basically suitable for ground-state and thermal properties, and is not well adapted to excited-state spectroscopy.

In parallel, the constrained-path~\cite{zhang1997constrained} or, more generally, the phaseless~\cite{zhang2003quantum} auxiliary-field QMC (AFQMC) approach has recently shown particular promise. By imposing a fixed-node like constraint~\cite{van1994fixed,ceperley1980ground} on branching random walks in Slater determinant space, this method can effectively control the sign or phase problem while maintaining systematically improvable accuracy. With appropriate trial wave functions, the approach can also target low-lying excited states~\cite{PhysRevLett.111.012502,bonnard2016constrained}. This method has been successfully applied to challenging molecular systems~\cite{mahajan2022selected,lee2020performance}, simple transition metal complexes~\cite{williams2020direct,rudshteyn2022calculation}, and solids~\cite{xu2024coexistence,motta2020ground}. Nevertheless, its broader applicability and development remain under active investigation~\cite{lee2022twenty,shi2021some,motta2018ab}, and applications to nuclear structure, where large discrete bases are required, are still in their infancy. In this work, unless otherwise specified, AFQMC refers exclusively to its phaseless formulation.

Like all constrained QMC methods, phaseless AFQMC is approximate, and its accuracy and efficiency are highly sensitive to the quality of trial wave function, which both defines the phase constraint and shapes the stochastic dynamics~\cite{zhang2003quantum}. Considerable effort has therefore focused on improving trial states, from single Slater determinants obtained via Hartree-Fock~\cite{lee2020utilizing}, DFT~\cite{zhang2003quantum}, or self-consistent calculations~\cite{qin2016coupling,qin2023self,he2019finite}, to multi-Slater~\cite{mahajan2022selected}, Jastrow-correlated~\cite{chang2016auxiliary}, and coupled-cluster forms~\cite{huggins2022unbiasing}. While these sophisticated trial states can improve performance, their implementation in AFQMC remains technically demanding and is often computationally expensive~\cite{xiao2025implementing,mahajan2020efficient}. 

Meanwhile, a different and largely unexplored avenue exists. Due to the overcomplete representation of Slater determinants, the AFQMC stochastic process is inherently non-unique. Yet, this non-uniqueness has received surprisingly little attention in AFQMC literature as a potential lever for improving performance. In contrast, this feature has been systematically exploited in Gaussian phase-space QMC~\cite{corney2006gaussian,corney2005gaussian} and was formalized as stochastic gauge freedom, allowing dynamically modifying the underlying stochastic process to suppress instabilities from heavy-tailed distributions~\cite{deuar2001stochastic,rousse2023simulations,deuar2002gauge} although optimal gauge selection remains nontrivial and challenging. Interestingly, AFQMC in a manifold of Slater determinants can be viewed as a special case of Gaussian phase-space QMC in a pure number-conserving Gaussian operator basis. This formal connection provides useful guidance and motivates the exploration of stochastic gauge analogues within the AFQMC framework to dynamically tune the stochastic process for improved sampling efficiency and accuracy. Therefore, apart from trial-state improvement to shape the underlying stochastic process, it is possible to introduce and develop such gauge-like transformations in AFQMC, in order to introduce additional flexibility and the potential for enhanced performance. To our knowledge, this possibility has not been systematically explored. This work addresses this gap by developing a new gauge-augmented AFQMC scheme. As a proof of concept, we investigate the approach with a mimic shell-model Hamiltonian, while highlighting the formalism differences and methodological insights that pave the way for extension to more complex systems in the future.

This paper is organized as follows. Section~\ref{Sec: AFQMC} briefly reviews the AFQMC formalism. Section~\ref{Sec: GAFQMC} introduces stochastic-gauge analogues within the AFQMC framework and formulates alternative stochastic schemes, which are analyzed and compared theoretically with the conventional approach. Section~\ref{Sec: results} presents numerical benchmarks of these schemes using a representative model Hamiltonian. Finally, conclusions and prospects are discussed in Section~\ref{Sec: con}.

\section{The AFQMC method}\label{Sec: AFQMC}

In this section, we introduce key notations and fundamental components of AFQMC and its implementation. More comprehensive reviews can be found in Refs.~\cite{lee2022twenty,shi2021some,motta2018ab}. We consider a general system of fermions interacting via a two-body interaction, described by a second-quantized Hamiltonian:
\begin{align} \label{Eq: ab Hamiltonian}
    \an H=\sum_{i \bar j}t_{i \bar j}\cra{c}_i\an c_{\bar j} + \frac{1}{2}\sum_{ik,\bar j \bar l}V_{ik,\bar j \bar l}\cra{c}_i\cra{c}_k\an{c}_{\bar l}\an c_{\bar j},
\end{align}
where $\cra{c}_i$ and $\an c_{\bar j}$ are fermionic creation and annihilation operators associated with a complete set of single-particle states $\ket{i}$ and their biorthogonal counterparts $\ket{\bar j}$, satisfying $\iner{\bar j}{i} = \delta_{ij}$. For simplicity, we consider a separable interaction of the form $V_{ik,\bar j \bar l}=-\lambda O_{i \bar j}O_{k \bar l}$, although extending the discussion to more general interactions is straightforward~\cite{koch2003reduced,lang1993monte}.

As in other projective QMC methods, AFQMC obtains the ground state through imaginary-time evolution:
\begin{align}
    \ket{\Psi_{\mathrm{gs}}} \propto \lim_{\tau \rightarrow \infty}e^{-\tau \an H} \ket{\Psi_0},
\end{align}
where $\ket{\Psi_0}$ is an arbitrary initial state with nonzero overlap with the exact ground state and is typically chosen to be the trial state $\ket{\Psi_T}$. In a path-integral formulation, this projection is carried out through a sequence of short time steps. Each small-time propagator is factorized using Trotter-Suzuki decomposition~\cite{trotter1959product} into one-body and two-body components, with the latter further decoupled via a Hubbard-Stratonovich transformation~\cite{hubbard1959calculation}. As a result, the full imaginary-time propagator can be reformulated as a high-dimensional path integral over auxiliary fields corresponding to one-body transformations, which transforms between Slater determinants via Thouless's theorem~\cite{thouless1960stability, thouless1961vibrational}. The resulting integral can then be evaluated stochastically using Monte Carlo sampling, thereby mapping the exact dynamics into a stochastic process over the Slater determinant manifold. 

However, direct sampling of this process typically suffers from poor efficiency and severe sign or phase problems. To mitigate these issues, an importance-sampling transformation is introduced in AFQMC~\cite{zhang2003quantum}. This transformation significantly improves convergence and numerical stability with a well-chosen trial wave function $\ket{\Psi_T}$. By employing a force bias in the importance-sampling transformation to minimize weight fluctuation~\cite{zhang2003quantum}, and using Ito calculus for stochastic integral~\cite{gardiner1985handbook}, the evolution of the single-particle orbitals $\ket{\varphi_k}$ in the walker state $\ket{\Psi^s_\tau}$, along with their associated weights $\Omega_\tau$, can be described by the following stochastic differential equations (SDEs):
\begin{align}\label{AFQMC}
 d\Omega/\Omega =& -d\tau \av{\an H}_{T,s}, \nonumber\\
 d\ket{\varphi_k} =& -d\tau (t - \lambda \av{\an O}_{T,s}O)\ket{\varphi_k} + dW\sqrt{\lambda} O \ket{\varphi_k},
\end{align}
where $\av{\cdot}_{T,s}$ denotes the mixed estimator between $\Psi_T$ and $\Psi^s_\tau$, and $dW$ is a Wiener increment satisfying $\mathbb{E}[dW] = 0$ and $dW_idW_j = d\tau \delta_{ij}$. In practice, the resulting stochastic process can be simulated either by direct integration of the above differential equations~\cite{kloeden1992stochastic} or via open-ended random walk algorithms~\cite{zhang1997constrained, zhang2003quantum}. In addition, population control and periodic orthonormalization of walker orbitals are employed to enhance numerical stability.  

In general, both the weights and orbitals are complex quantities. Without any constraints, the weights are free to wander over the complex plane during the stochastic evolution, leading to a severe phase problem. To mitigate this issue, the phaseless approximation is proposed with an approximate trial wave function~\cite{zhang2003quantum}. This involves a positive-weight approximation and a cosine projection of the rotation angle $\Omega_i' = \left|\Omega_i\right|\times {\rm max}\left( 0, \cos(\Delta \theta)\right)$ with
\begin{align}
    \Delta \theta =  \arg\left(\frac{\iner{\Psi_T}{\Psi^s_{i+1}}}{\iner{\Psi_T}{\Psi^s_i}}\right).
\end{align}
Importantly, in the exact limit where the trial wave function coincides with the true ground state, the local energy $\av{\an H}_{T,s}$ in Eq.~(\ref{AFQMC}) becomes real and the weights remain positive, thereby explicitly eliminating the phase problem. This property provides a pathway for improving the accuracy of the phase constraint through systematic improvement of the trial wave function, which remains a central focus of ongoing developments in AFQMC~\cite{shi2021some}. In the next section, we propose alternative stochastic schemes that exploit the intrinsic non-uniqueness of the underlying stochastic process.

\section{Gauge AFQMC} \label{Sec: GAFQMC}
AFQMC maps the exact many-body dynamics onto a stochastic process over the manifold of Slater determinants, or more generally, quasiparticle vacua~\cite{shi2017many,vitali2024monte,juillet2017phaseless}. The overcomplete nature of this representation implies that the associated stochastic dynamics are inherently non-unique. This non-uniqueness offers a degree of flexibility that can be exploited for optimization, analogous to stochastic gauge in phase-space QMC methods~\cite{corney2006gaussian,deuar2002gauge}. In the context of AFQMC, similar gauge strategies can be identified. For instance, different choices of Hubbard-Stratonovich transformations, arising from non-unique decompositions of Hermitian operators, lead to distinct stochastic representations and can be interpreted as diffusion gauge in the language of Gaussian phase-space QMC~\cite{corney2006gaussian}. In this section, we adopt the terminology of Gaussian representations and formulate several stochastic gauge analogues within the AFQMC framework.   
\subsection{Drift gauge}\label{sec: drift}
In Gaussian phase-space QMC, drift gauge enables a tunable trade-off between the drift in phase-space variables and the diffusion in the weight~\cite{corney2006gaussian}. An analogous mechanism can be introduced in AFQMC via a generalized background shift. Specifically, the Hamiltonian of Eq.~\eqref{Eq: ab Hamiltonian} can be reformulated as
\begin{align}
    \an H = \frac{\lambda }{2}g_s^2+\an T + \lambda g_s\an O - \frac{\lambda }{2}(\an O + g_s)^2,
\end{align}
where the one-body matrix $T = t + \lambda O^2/2$ and $g_s$ is a local, trajectory-dependent shift. This formulation closely resembles the mean-field subtraction techniques employed to improve the phase constraint in AFQMC~\cite{shi2013symmetry,al2006auxiliary}. Based on this modified Hamiltonian, the importance-sampling dynamics of the weights $\Omega$ and walkers $\ket{\Psi^s}$ follow from the standard derivation with the operator substitutions:
\begin{align}
    \an T \longrightarrow \an T + \lambda g_s \an O, \quad
    \an O^2 \longrightarrow (\an O + g_s)^2.
\end{align}
The scalar term is absorbed into the weight evolution. The resulting Ito SDEs, now augmented by the parameter $g_s$, take the form:
\begin{align}
    d\Omega/\Omega =& -d\tau \av{\an H}_{T,s} + dW\sqrt{\lambda}(\av{\an O}_{T,s} + g_s), \nonumber\\
 d\ket{\varphi_k} =& -d\tau (t + \lambda g_s O)\ket{\varphi_k}+ dW\sqrt{\lambda} O \ket{\varphi_k}.
\end{align}
Notably, there remains a degree of freedom in whether the stochastic term proportional to $g_s dW$ is incorporated into the walker evolution. 

As shown above, the shift parameter $g_s$ acts as a drift gauge, trading off between contributions to the walker orbitals and the weight fluctuations, analogous to its role in Gaussian QMC. To minimize weight diffusion, one may choose $g_s = - \av{\an O}_{T,s}$, which exactly recovers the conventional AFQMC scheme described in Eq.~(\ref{AFQMC}). In this sense, the commonly used force bias in importance-sampling AFQMC corresponds to a particular choice of drift gauge that minimizes weight diffusion. The resulting drift term subtracts only the direct (Hartree) contribution of the interaction. As shown in the following subsections, it is possible to incorporate both direct and exchange effects to the drift term while controlling weight fluctuations. Moreover, the conventional choice of drift gauge is not always optimal and alternative choices of $g_s$ may yield improved performance in certain cases, albeit at the cost of increased variance in the weight distribution.

\subsection{Fermi gauge}\label{sec: fermi}
In Gaussian phase-space QMC, additional gauge freedoms arise in fermionic systems, termed Fermi gauge, which uses Fermi statistics (e.g., $\an a^2 =0$) to add zero terms that generate different but equivalent SDEs~\cite{corney2006gaussian}. A similar flexibility exists in AFQMC: terms that annihilate vacuum state or many-body walker (e.g., $\cra c_i \ket{0_c} = 0$) may be added to the dynamics, allowing for different stochastic formulations that preserve the exact imaginary-time evolution. These zero terms depend explicitly on the structure of stochastic walker, which can be either a Slater determinant or a Bogoliubov quasiparticle vacuum~\cite{shi2017many,vitali2024monte,juillet2017phaseless}. In this work, we focus on the formulation using Slater determinant walkers. Extension to the Bogoliubov quasiparticle case can be carried out in a similar manner.

To be specific, we have
\begin{align}
    0 = g_{\bar h_ih_j}\an c_{\bar h_i} \cra c_{h_j} \ket{\Psi^s} = (g_{\bar h_i h_j} \delta_{ij} - g_{\bar h_i h_j} \cra c_{h_j} \an c_{\bar h_i})\ket{\Psi^s},
\end{align}
where $\ket{\Psi^s} = \prod_k \cra c_{h_k} \ket{-}$ is a stochastic Slater determinant walker in AFQMC. The hole states $\ket{h_i}$ and the particle state $\ket{p_i}$ span the one-body Hilbert space, with the closure relation given by $1 = \sum_i\ket{p_i}\bra{\bar p_i} + \sum_i\ket{h_i}\bra{\bar h_i}$, where $\bra{\bar h_i}$ and $\bra{\bar p_i}$ denote the biorthogonal components of $\ket{h_i}$ and $\ket{p_i}$, respectively, to account for the non-orthogonality arising from the non-unitary stochastic projection. Thus, the hole-hole ($hh$) components of the one-body operators serve as zero terms in AFQMC, i.e., $g_{\bar hh}\an c_{\bar h} \cra c_{h} \sim 0$. With these zero terms, the one-body components of the Hamiltonian can be modified, thereby altering the drift term of the walker. Similarly, for the two-body operators, we have
\begin{align}
    0\sim& g_{p_i h_j \bar h_k \bar h_l}\cra c_{p_i} \an c_{\bar h_l} \an c_{\bar h_k} \cra c_{h_j} \nonumber \\
    =& g_{p_i h_j \bar h_k \bar h_l} (\cra c_{p_i} \cra c_{h_j} \an c_{\bar h_l} \an c_{\bar h_k} + \cra c_{p_i}\an c_{\bar h_l}\delta_{jk} - \cra c_{p_i}\an c_{\bar h_k}\delta_{jl}).
\end{align}
These terms can modify the one- and two-body components of the Hamiltonian, leading to different drift
and diffusion terms in the resulting SDEs without altering the underlying dynamics. Notably, the choice of the biorthogonal basis is not unique and can itself be treated as a gauge degree of freedom in this formulation. 

To provide a concrete example of an alternative stochastic scheme, we exploit these zero terms in the natural-orbital basis of a reduced one-body density matrix (ROBDM) and simplify the two-body interaction to particle($p$)-hole($h$) channels using Wick's theorem. The ROBDM is defined via a mixed estimator between the trial wave function $\ket{\Psi_T}$ and the stochastic Slater determinant walker $\ket{\Psi^s}$ as
\begin{align}
    \rho_{ij} = \frac{\mat{\Psi_T}{\cra c_j \an c_i}{\Psi^s}}{\iner{\Psi_T}{\Psi^s}} = \av{\cra c_j \an c_i}_{T,s}.
\end{align}
It can be verified that $\rho^2 = \rho$, regardless of whether the trial state is a single Slater determinant or a correlated linear superposition. This property have been exploited in the construction of self-consistent trial wave functions~\cite{qin2023self}. Accordingly, $\rho$ is a generally non-Hermitian projection matrix, and its natural orbitals form a biorthogonal basis in the one-body Hilbert space. The single-particle states of $\ket{\Psi^s}$ correspond to the right eigenstates of $\rho$ and span the hole space. The Hamiltonian can then be normal-ordered with respect to this vacuum using contractions defined by the ROBDM, yielding
\begin{align}
    \an H = \av{\an H}_w + \sum_{ij}h_{i j}:\cra c_i \an c_{ j}: + \sum_{ikjl}\frac{1}{2}V_{ik,jl}:\cra c_i \cra c_k \an c_{ l} \an c_{ j}:,
\end{align}
where $\av{\an H}_w = {\rm tr}(t\rho) + \frac{1}{2}{\rm tr}(\overline{V}(\rho)\rho)$ is the Wick's contraction term and $ h = t + \overline{V}(\rho)$ takes the form of the standard Hartree-Fock potential with $\overline{V}(\rho)_{i j} = \sum_{k l}(V_{ik,  j  l} - V_{ik,  l  j})\rho_{ lk}$.
Note that $\av{\an H}_w$ is not the mixed estimator of the Hamiltonian when $\Psi_T$ is not a single Slater determinant. The normal-ordering interaction term can then be decoupled via Hubbard-Stratonovich transformation, giving
\begin{align}
    :e^{-\Delta \tau\an V}:\ket{\Psi^s} =& \int d\sigma G(\sigma):e^{\sigma \sqrt{\Delta \tau}\sqrt{\lambda}\an O}:\ket{\Psi^s} \nonumber \\
    =& \int d\sigma G(\sigma)e^{\sigma \sqrt{\Delta \tau}\sqrt{\lambda}\an O^{ph}}\ket{\Psi^s}, 
\end{align}
where $G(\sigma)$ is the standard Gaussian distribution, $O^{ph} = (1-\rho)O\rho$, and $(O^{ph})^2=0$. The weight terms in the importance-sampling transformation are then iterated as
\begin{align}
    \Omega_{i+1} = \Omega_i e^{-\Delta \tau \av{\an H}_w} \av{e^{-\Delta \tau \an h^{ph} + \Delta W \sqrt{\lambda}\an O^{ph}}}_{T,s},
\end{align}
where $\Delta W$ is the Wiener increment. Using relations $\av{\an h^{ph}}_{T,s}=0$, $\av{\an O^{ph}}_{T,s} = 0$, and the convention $\Delta W ^2 = \Delta \tau$, we obtain the standard local energy expression:
\begin{align}
    \Omega_{i+1} = &\Omega_i \exp\left(-\Delta \tau \av{\an H}_w -\Delta \tau \av{\an V^{pphh}}_{T,s}\right)\nonumber\\
    =&\Omega_i \exp\left(-\Delta \tau \av{\an H}_{T,s}\right),
\end{align}
recovering the conventional AFQMC weight update~\cite{zhang2003quantum}. Introducing Ito integral~\cite{gardiner1985handbook}, the continuous-time stochastic evolution reads
\begin{align}
    \Omega(\tau) =& \Omega(0)\exp\left(-\int_0^\tau dt \av{\an H}_{T,s} \right), \nonumber \\
    \ket{\Psi^s(\tau)} =& \exp\left (-\int_0^\tau dt \an h + \sqrt{\lambda}\int_0^\tau dW\an O^{ph}\right)\ket{\Psi^s(0)}.
\end{align}
Note that we modify the drift term of the walker to the standard mean-field potential using the gauges $g_{hh}$. 

In particular, the resulting Ito SDEs for the single-particle orbitals $\ket{\varphi_k}$ are given by, using Ito calculus and $(O^{ph})^2 = 0$,
\begin{align}
    d\ket{\varphi_k} = -d\tau h(\rho)\ket{\varphi_k} + dW \sqrt{\lambda}(1-\rho)O\ket{\varphi_k}.
\end{align}
The drift terms thus correspond exactly to the standard mean-field Hartree-Fock (HF) potential. Hence, the exact imaginary-time dynamics can be interpreted as a time-dependent HF like evolution augmented by fluctuations in the $ph$ channels. 

Compared with the conventional AFQMC scheme [Eq.(\ref{AFQMC})] which subtracts only the direct term of the interaction ($-\lambda \av{\an O}_{T,s}O$) in the drift dynamics, the present mean-field-like stochastic formulation preserves the same local energy structure for weight propagation while introducing a modified drift term that incorporates both Hartree ($-\lambda \av{\an O}_{T,s}O$) and Fock ($\lambda O\rho O$) contributions. The diffusion is minimized and restricted to the $ph$ channels defined through the natural orbitals of the ROBDM.

In analogy to the background subtraction technique employed in the phaseless approximation to reduce phase fluctuations~\cite{shi2013symmetry,al2006auxiliary}, Fermi gauges subtract more of the interaction in the drift, reducing the magnitude of stochastic fluctuations. This reduction may lead to a smaller phaseless bias, particularly when cosine-projection rejections become significant. The mean-field scheme is less sensitive to the cosine projection since the rotation angle $\Delta \theta$ scales as $\mathcal{O}(\Delta \tau)$ rather than $\mathcal{O}(\sqrt{\Delta \tau})$. 

Similar mean-field structures have been observed and benchmarked in earlier works~\cite{juillet2002exact, juillet2007sign}, where they were formulated using a single Slater determinant trial state. As shown in the present work, the formulation can be naturally extended to more correlated trial states via the generalized Wick's theorem. Such correlated trial states are known to improve stochastic guidance and sampling efficiency. In the following, we refer to this approach as mean-field AFQMC (MF-AFQMC) and illustrate, in the next section, the numerical improvements arising from these modifications.

\subsection{Symmetry gauge}
Symmetry-breaking and symmetry-restoring techniques have long been employed in nuclear physics to extend descriptions beyond the mean-field picture~\cite{TB:ring2004nuclear}. In AFQMC, these methods have been applied to improve the stochastic scheme by incorporating a symmetry-projected trial wave function and a symmetry-breaking Hubbard-Stratonovich (SB-HS) transformation~\cite{shi2013symmetry}. An enlarged, overcomplete basis generated by symmetry breaking is generally expected to improve stochastic sampling efficiency since each walker carries more information. For instance, a U(1) symmetry-breaking walker can capture pairing correlations that would otherwise require a large superposition of Slater determinants. However, the SB-HS transformation may also lead to increased statistical fluctuations~\cite{shi2013symmetry}. This arises because different symmetry sectors may mix during propagation, resulting in contamination from unwanted symmetry channels. To retain the advantages of SB-HS transformations while controlling sector contamination, we propose a symmetry-projection sampling scheme that monitors and suppresses unwanted symmetry components during propagation. The mechanism is described below and illustrated for parity symmetry breaking in the next section.

In the language of group theory, the symmetry projection operator $S^{(\nu)}$ for a $n_\nu$-dimensional irreducible representation $\nu$ of a finite symmetry group $G$ can be formally expressed as
\begin{align}
    \mathcal{S}^{(\nu)} = \frac{n_{\nu}}{g}\sum_{R \in G} \chi^{\nu}(R)^* \mathcal{P}(R),
\end{align}
where $g$ is the order of the group, $\mathcal{P}$ is a general (not necessarily irreducible) representation, and $\chi^{\nu}$ is the character of the unitary irreducible representation $\nu$. The projection operator can be rewritten in a probabilistic form as
\begin{align}
    \mathcal{S}^{(\nu)} = \sum_{ G} p(R) \mathcal{U}(R),
\end{align}
where $p(R) = |\chi^\nu(R)|^2/g$ and $\mathcal{U}(R) = n_\nu \mathcal{P}(R)/\chi^\nu(R)$. Using the orthogonality relations, one can verify that $0\leq p(R) \leq 1$ and $\sum p(R) = 1$. These results can be extended to the continuous case using the Haar measure~\cite{hamermesh2012group}.

With the explicit expressions above, we apply the symmetry projection operator to a symmetry-breaking AFQMC walker within the importance-sampling framework as follows:
\begin{align}
    \mathcal{S}^{(\nu)}\frac{\ket{\Psi^s_i}}{\iner{\Psi_T}{\Psi^s_i}} = \int da p(a) \mathcal{U}(a)\frac{\ket{\Psi^s_i}}{\iner{\Psi_T}{\Psi^s_i}}.
\end{align}
When $\mathcal{U}(a)$ is a one-body transformation, the transformed walker remains a Slater determinant via Thouless's theorem. Thus, the stochastic process remains confined to the Slater determinant manifold. The parameter $a$ can be treated as an additional stochastic variable sampled from $p(a)$. The resulting symmetry-projected importance-sampling propagator then takes the form
\begin{align}
    \int d\sigma da G(\sigma)p(a) \iner{\Psi_T}{\Psi^{s,a}_{i+1}}\an B(\sigma)\mathcal{U}(a) \frac{1}{\iner{\Psi_T}{\Psi^s_i}}  ,
\end{align}
where $\an B(\sigma)$ is the stochastic propagator arising from the HS transformation, $\ket{\Psi^{s,a}_{i+1}} = \an B(\sigma)\mathcal{U}(a)\ket{\Psi^s_i} = \an B(\sigma)\ket{\Psi^{s,a}_i}$, and $p(a)$ is the probability distribution of the symmetry parameter $a$. In particular, for abelian symmetries, such as $Z_2$ and $U(1)$ symmetries, we may choose the trial wave function to transform as a one-dimensional irreducible representation. In this case, $\bra{\Psi_T}\mathcal{U}(a) = \bra{\Psi_T}$ and $\iner{\Psi_T}{\Psi^s_i} = \iner{\Psi_T}{\Psi^{s,a}_i}$. The weight evolution reduces to the standard local energy form,
\begin{align}
    \Omega_{i+1} = \Omega_i \exp(-\Delta \tau \av{\an H}_{\Psi_T, \Psi^{s,a}_i}).
\end{align}
Notably, for these abelian cases, the symmetry projection does not introduce additional phase factors that could exacerbate the sign or phase problems. In practice, the symmetry projection need not be applied at every iteration; instead, it may be implemented adaptively when the effects of symmetry-breaking contamination become significant, as indicated by deviations between the sampled observables and their estimated values. We refer to this approach as symmetry-projection sampling in the following discussion.

\section{Illustration} \label{Sec: results}
In Sec.~III, we formulate a generalized framework that recovers conventional phaseless AFQMC as a particular choice of drift gauge while introducing additional gauge freedoms that are theoretically appealing and may enable further improvements. The conventional gauge has already been successfully applied to realistic nuclear shell-model Hamiltonians with phenomenological interactions in the presence of a sign problem~\cite{PhysRevLett.111.012502}. The generalized formulation preserves the original importance-sampling structure of AFQMC and can therefore be implemented with minimal modification of existing algorithms. Thus, the present formulation retains the established applicability of the phaseless AFQMC to realistic systems, while providing a systematic extension that may reduce bias and statistical fluctuations.

In this section, as a numerical proof of concept, we employ the Lipkin-Meshkov-Glick (LMG) model~\cite{LMG1}. Although highly simplified, the LMG model captures essential features of interacting many-body systems, including collective correlations, symmetry breaking, and quantum phase transitions~\cite{LMG4}. At the same time, its exact solvability and modest Hilbert-space dimension allow for controlled benchmarking of systematic biases and statistical fluctuations. These properties make it an ideal testing ground for isolating and illustrating the impact of different gauge choices and assessing their effect on sampling efficiency and accuracy before proceeding to more realistic shell model applications.

\begin{figure}[b]
    \centering
    \includegraphics[width=0.45\textwidth]{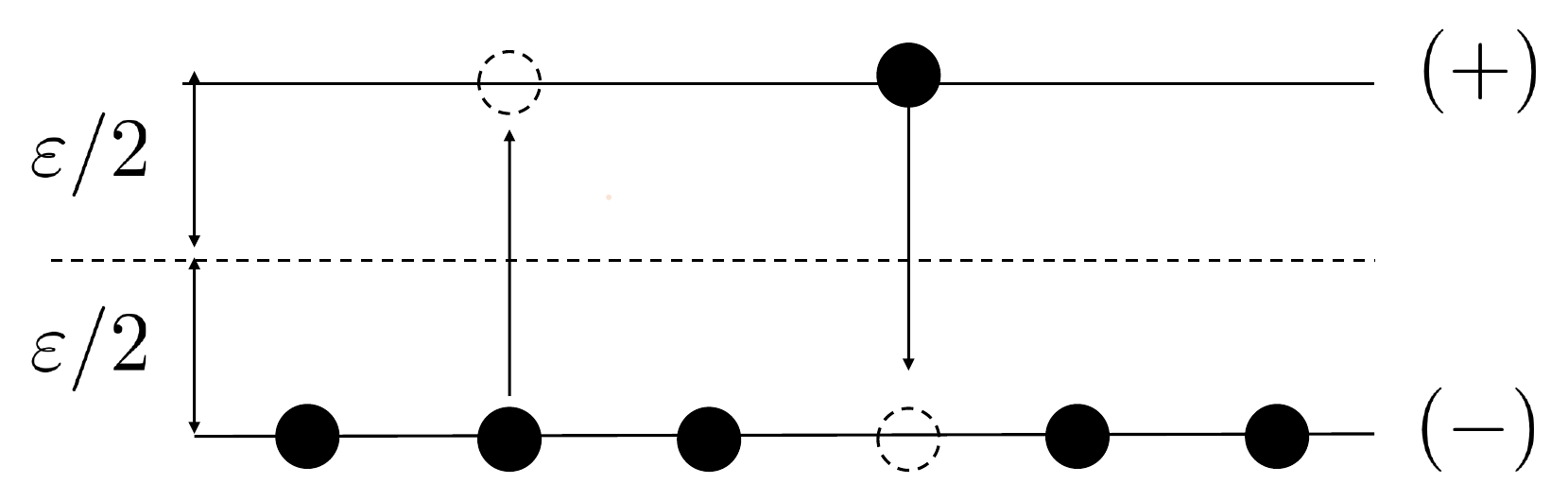}
    \caption{A configuration of the LMG model with $N = 6$ particles, in which transitions can only occur directly upward or downward due to the monopole-monopole interaction.}
    \label{fig:LMG}
\end{figure}
The LMG model describes a system of $N$ particles distributed across two $N$-fold degenerate energy levels separated by an energy gap $\varepsilon$. The corresponding $N$-fermion Hamiltonian is given by
\begin{align}
    \an H=\frac{\varepsilon}{2}\sum_{p\sigma}\sigma\cra{a}_{p\sigma}\an a_{p\sigma} 
    &-\frac{V}{2}\sum_{p,p^\prime}(\cra{a}_{p+}\cra{a}_{p^\prime+}\an a_{p^\prime-} \an a_{p-}+h.c.) \nonumber \\
    &-\frac{W}{2}\sum_{p,p^\prime}(\cra{a}_{p-}\cra{a}_{p^\prime+}\an a_{p^\prime-} \an a_{p+}+h.c.),
\end{align}
where $\cra a_{p\sigma}$ and $\an a_{p\sigma}$ are the fermionic creation and annihilation operators for level $\sigma \in \{+,-\}$ (denoting the upper and lower levels) and degeneracy index $p$. The $V$-term interaction corresponds to pairwise scattering between particles in the same level, while the $W$-term interaction involves one particle moving up and another moving down. Introducing the collective quasispin operators $\an J_z = \frac{1}{2} \sum_{p\sigma}\sigma\cra a_{p\sigma} \an a_{p\sigma}$, $\an J_x = \frac{1}{2}\sum_{p\sigma}\cra a_{p\sigma}\an a_{p-\sigma}$, and $\an J_y = \frac{1}{2i}\sum_{p\sigma}\sigma\cra a_{p\sigma}\an a_{p-\sigma}$, the Hamiltonian can be rewritten as
\begin{align}
    \an H=& \varepsilon \an J_z - \an J_x^2(V + W) - \an J_y^2(W - V) \nonumber \\
    =& \varepsilon \an J_z - \an J_x^2\alpha_x -  \an J_y^2\alpha_y,
\end{align}
where $\alpha_x = V + W$ and $\alpha_y = W - V$. The LMG model is solvable due to its underlying SU(2) algebraic structure. The Hamiltonian is also invariant under $\pi$-rotation around the $z$-axis, which can be interpreted as a parity symmetry. The associated parity operator can be defined as $\an \Pi = e^{i\pi(\an J_z + N/2)}$, for which the unperturbed ground state $\ket{{\rm u.g.s}} = \prod_p \cra c_{p-}\ket{-}$ has positive parity. The Hartree-Fock (HF) approximation to the ground state can be obtained by an uniform rotation in the quasispin space: $\ket{{\rm HF}} = e^{-i\theta \an J_y}\ket{{\rm u.g.s}}$, where the rotation angle is determined by minimizing the HF energy. For $\chi = (N-1)\alpha_x/\varepsilon < 1$, one finds $\theta =0$, corresponding to a spherical HF state with definite parity. For $\chi > 1$, $\theta \neq 0$, corresponding to a deformed HF state with parity-symmetry breaking. The symmetry can be restored by the projection operators $\an P_{\pm} = \frac{1}{2}(1 \pm \an \Pi)$, which, in most case, improves the accuracy of mean-field approximations.

\begin{figure}[t]
\centering
\includegraphics[width=0.4\textwidth]{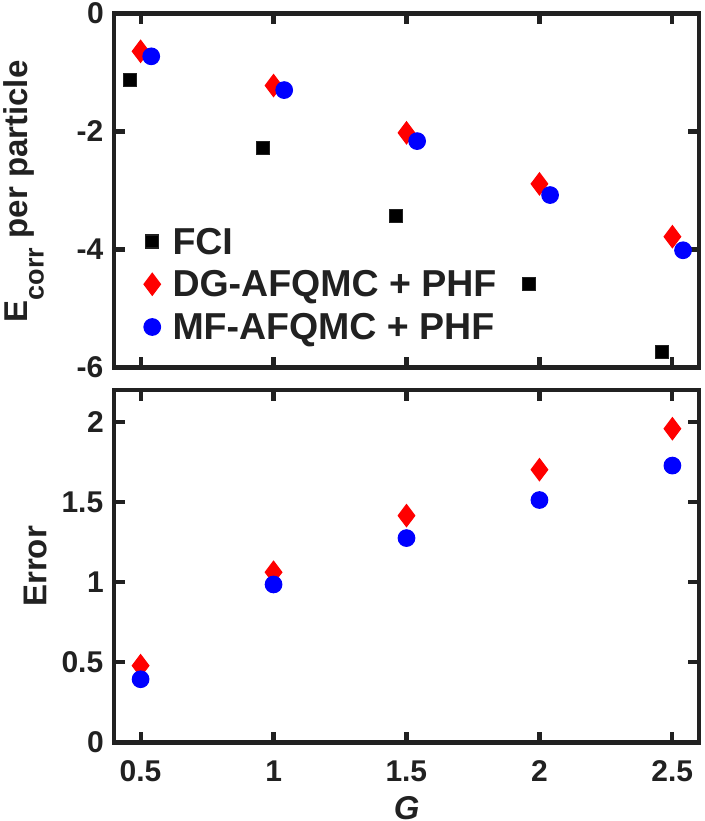}
    \caption{Correlation energy of the LMG model for $N = 10$ and $\alpha_y = 2\alpha_x =2 G\varepsilon$, where $G$ denotes the dimensionless coupling strength. The statistical uncertainty of the total correlation energy is on the order of $1\%$ and is too small to be visible. Top: correlation energy per particle as a function of $G$. Bottom: deviation of the QMC results from the exact FCI values obtained by exact diagonalization of the LMG Hamiltonian. We compare the performance of MF-AFQMC with that of conventional AFQMC (denoted DG-AFQMC), with both QMC simulations guided by a parity-projected HF (PHF) trial state.}
    \label{fig:LMG_1_2}
\end{figure}

\begin{figure}[t]
\centering
\includegraphics[width=0.4\textwidth]{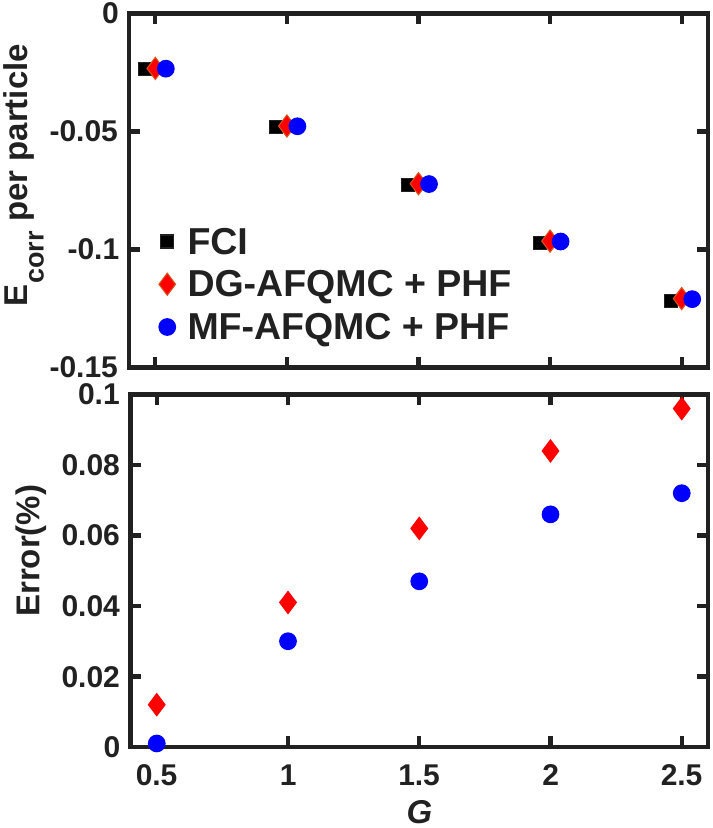}
    \caption{Correlation energy of the LMG model for $N = 10$ and $\alpha_x = 2\alpha_y =2 G\varepsilon$, where $G$ denotes the dimensionless coupling strength. The statistical uncertainty of the total correlation energy is on the order of $0.1\%$ and is too small to be visible. Top: correlation energy per particle as a function of $G$. Bottom: deviation of the QMC results from the exact FCI values obtained by exact diagonalization of the LMG Hamiltonian. We compare the performance of MF-AFQMC with that of conventional AFQMC (denoted DG-AFQMC), with both QMC simulations guided by a parity-projected HF (PHF) trial state.}
    \label{fig:LMG_2_1}
\end{figure}

In our AFQMC simulation, we employ a direct Hubbard-Stratonovich transformation to linearize the quadratic interacting terms, as described in Sec.~\ref{Sec: AFQMC}. This procedure generates parity-breaking Slater determinant walkers. We begin by comparing the performance of the mean-field AFQMC (MF-AFQMC) method introduced in Sec.~III.B with the conventional approach, referred to as drift gauge AFQMC (DG-AFQMC) since it can be interpreted within the present framework as a drift gauge transformation presented in Sec.~III.A. Both methods use parity-projected HF solutions as the trial wave functions, i.e., $\ket{\Psi_T} = \an P_+ \ket{\rm HF}$. The correlation energies, computed using the mixed estimator, are benchmarked against exact full configuration interaction (FCI) results. The results are presented in Figs.~\ref{fig:LMG_1_2} and \ref{fig:LMG_2_1}.

Fig.~\ref{fig:LMG_1_2} shows the correlation energy as a function of the dimensionless coupling strength $G$ for the LMG model with $\alpha_y/\varepsilon = 2\alpha_x/\varepsilon = 2G$ and $N = 10$. The top panel displays the correlation energies per particle, while the bottom panel presents the deviations of the QMC results from the FCI solutions. In this regime, the real HF approximation fails to capture the dominant correlations, particularly at strong coupling. The AFQMC simulations suffer from a phase problem over long-time simulation, which is controlled by the standard phaseless approximation. The accuracy of this approximation depends sensitively on the quality of the trial wave function, and a noticeable bias emerges when the trial state is inadequate, as illustrated in Fig.~\ref{fig:LMG_1_2}. Nevertheless, we emphasize that the results can be substantially improved by employing a higher-quality trial state, as shown in Fig.~\ref{fig:LMG_2_1} for the LMG model with $\alpha_x/\varepsilon = 2\alpha_y/\varepsilon =2 G$, which is related to the previous Hamiltonian through a rotation. In this case, the real HF approximation provides a significantly improved description of the ground state, with correlation energies at the $\~10\%$ level. With these improved trial wave functions, AFQMC reproduces the exact results with deviations at the $0.1\%$ level. 

In both cases the sign problem is mild and the phaseless bias remains small when a reasonable trial state is employed, even at the HF level. As discussed in Sec.~\ref{sec: fermi}, the MF-AFQMC scheme reduces stochastic diffusion by incorporating both Hartree and Fock contributions into the drift term, thereby restricting orbitals fluctuations to particle-hole channels. This reduction is expected to mitigate phaseless bias, particularly when cosine-projection rejections become significant. The observed improvement from MF-AFQMC is consistent with this expectation, although the effect is modest due to the mild sign problem in the LMG model. We therefore expect the advantages of the mean-field scheme to become more pronounced in more complex systems with stronger sign problems and larger phase fluctuations, which will be explored in future work.

\begin{figure}[t]
\centering
\includegraphics[width=0.45\textwidth]{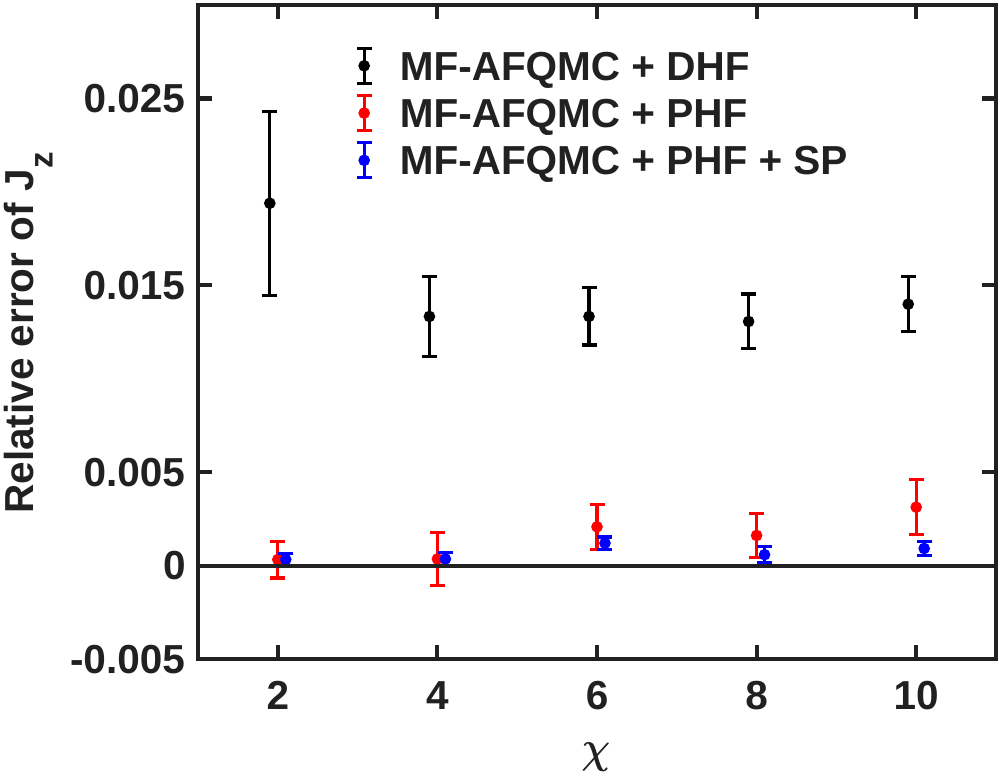}
    \caption{Relative error of the angular momentum observable $J_z$ in the LMG model with $N=10$ and $W = 0$ across various interaction strengths $\chi = (N-1)V/\varepsilon$. The relative error is defined as $\left| (\av{J_z}_{QMC} - \av{J_z}_{\rm FCI})/\av{J_z}_{\rm FCI} \right|$, and the error bars denote the QMC statistical uncertainty. Results are shown for MF-AFQMC simulations guided by a deformed HF (DHF) trial state and a parity-projected HF (PHF) trial state with and without symmetry-projection sampling (SP).}
    \label{fig:SG}
\end{figure}
Next, to illustrate the role of the symmetry gauge, we consider the LMG model with $W = 0$ (i.e., $\alpha_x = - \alpha_y$). In this regime, AFQMC does not encounter a phase problem in the attractive interaction region. The dominant source of systematic errors arises instead from numerical constraints, such as population control, introduced to enhance sampling efficiency and stability. Although parity symmetry is conserved by the Hamiltonian, it is restored only statistically within the stochastic evolution, and suppressing symmetry contamination generally requires a substantially larger sample size. To improve sampling efficiency, we introduce symmetry-projection sampling. To isolate its effect, we evaluate a noncommuting observable, $J_z$, using an exact estimator with a deliberately limited number of samples ($2\times10^4$), thereby amplifying statistical fluctuations. Figure~\ref{fig:SG} shows the deviations of the QMC estimators from exact FCI results over a range of interaction strengths and stochastic schemes. We compare three cases: MF-AFQMC guided by a deformed HF (DHF) trial state, MF-AFQMC guided by a projected HF (PHF) trial state, and MF-AFQMC combining the PHF trial state with symmetry-projection sampling (PHF+SP).

Across all interaction strengths considered, MF-AFQMC reproduces exact solutions with relative errors on the order of $1\%$. The DHF-guided simulations exhibit significantly larger bias and fluctuations, particularly near the critical point $\chi = 1$, where the HF approximation fails to capture the dominant correlations. Restoring parity symmetry at the trial-state level significantly improves the reference quality, leading to substantial reductions in both systematic bias and statistical fluctuations, as seen in the PHF-guided results. The improved statistical behavior with the PHF trial wave function can also be attributed to its symmetry-restored structure. In AFQMC simulations using an exact estimator, statistical weights depend on overlaps between walker states and the trial wave function. When using a symmetry-broken trial state such as DHF, components from orthogonal symmetry sectors may contribute, leading to contamination from excited states and enhanced statistical noise. In contrast, the PHF wave function projects explicitly the walker ensemble onto the target symmetry sector, suppressing such unwanted contributions and improving sampling efficiency, particularly for symmetry-sensitive observables such as $J_z$.

To further control symmetry contamination during stochastic evolution, we incorporate symmetry-projection sampling on top of the PHF-guided simulation. Due to the inherently symmetry-breaking nature of the HS transformation, the stochastic propagator inevitably mixes components from different parity sectors during Monte Carlo sampling, which may lead to increased statistical fluctuations in measured observables. Symmetry-projection sampling monitors the cumulative influence of symmetry-breaking components and applies symmetry projection adaptively when this influence exceeds a predefined threshold. For the parity symmetry considered here, the projection operator is sampled through a Bernoulli process. Moreover, by adopting a PHF trial wave function with fixed parity, we avoid any additional sign problem that could otherwise arise from symmetry-projection sampling. As shown in Fig.~\ref{fig:SG}, the PHF $+$ SP scheme further reduces statistical uncertainty relative to PHF alone. These results demonstrate that while symmetry restoration at the trial-state level already enhances AFQMC performance, combining it with symmetry-projection sampling during stochastic evolution provides a more effective and efficient strategy for controlling symmetry contamination. This approach should be particularly beneficial for observables that are sensitive to symmetry mixing, such as those in open-shell nuclei with strong pairing correlations, where unphysical mixing between different particle-number sectors during propagation may introduce larger fluctuations.

\section{Conclusion}\label{Sec: con}
In this work, we have proposed alternative QMC schemes by exploiting the stochastic gauge freedom originally developed in Gaussian phase-space QMC and incorporating it into the importance-sampling AFQMC framework with the phaseless constraint. In particular, we reinterpret the conventional force bias as a drift gauge, which mediates a trade-off between weight diffusion and walker drift in AFQMC dynamics. We further explore Fermi gauges in the natural orbital basis defined via the reduced one-body density matrix from a mixed estimator between the trial state and the stochastic walker. This leads to a mean-field-guided stochastic evolution in which Hartree and Fock contributions are incorporated directly into the drift term, while residual fluctuations are restricted to particle-hole excitation channels. In addition, we develop a symmetry-projection sampling scheme designed to mitigate symmetry contamination induced by the symmetry-breaking Hubbard-Stratonovich transformation. Together, these developments substantially expand the flexibility of the current AFQMC framework. Our numerical benchmarks demonstrate that the additional gauge freedom can yield comparable or even superior performance with higher accuracy and reduced fluctuations. While the gains are modest for the mild LMG model considered in this work, the formal structure introduced here establishes a systematic route for optimizing stochastic dynamics in more challenging many-body systems with stronger sign or phase instabilities, including many previously studied with conventional phaseless AFQMC method. Indeed, previous phaseless AFQMC calculations in realistic nuclear shell model applications have already achieved sub-MeV accuracy despite significant sign problems~\cite{PhysRevLett.111.012502}, suggesting substantial room for further improvement through refined control of stochastic gauges and continued methodological development of phaseless AFQMC. We expect that further benchmarks and realistic applications will clarify the full potential of this gauged-augmented perspective.

\section*{Acknowledgments}
The author thanks Haozhao Liang for his guidance. Discussions with Yinu Zhang are greatly appreciated. This research was supported by Forefront Physics and Mathematics Program to Drive Transformation (FoPM), a World-leading Innovative Graduate Study (WINGS) Program, the University of Tokyo.

\bibliography{bibliography}
\end{document}